\input epsf
\documentstyle[11pt,paspconf]{article}
\begin{document}

\title{Future Cosmic Microwave Background Experiments}
\author{Mark Halpern and Douglas Scott}
\affil{Department of Physics and Astronomy, University of British Columbia,
          Vancouver, B.C. V6T 1Z1\ \  Canada}

\begin{abstract}
We summarise some aspects of experiments currently being built or planned,
and indulge in wild speculation about possibilities on the more distant
horizon.
\end{abstract}

\keywords{Cosmic Microwave Background, experiments, speculation}

\section{Prologue}

The satellite missions {\sl MAP} and {\sl Planck} dominate any view of future
measurements of the anisotropy of the Cosmic Microwave Background.  We
will attempt to look beyond and around those two experiments and see
what sorts of physical questions other future projects might
address.

The reader has several advantages over the authors which we will not
try to counter.  First, many of the experiments which are in the near
term future for us will be in the present or past for the reader, so
we do not focus on evaluating detailed {\it anticipated\/} technical
capabilities for a short list of such experiments.  Readers who wish to
pursue that approach might start at
{\tt http://www.astro.ubc.ca/people/scott/cmb.html},
or other similar web-pages for up to date information and links.

Second, readers of the rest this volume will be in a better position than we
are to make informed judgements about the ideal strategies for
measuring, avoiding or understanding foreground sources.  Therefore,
even though we think that this aspect of anisotropy  will be an
increasingly important and sophisticated part of the field, we have
not put much emphasis on it here.  As a crude aid to understanding how
well future experiments are equipped to cope with foreground sources
we have included a column giving the number of independent frequency
channels for each experiment listed in Table~1.

A view of the present situation, indicated in Figure~1 (see Smoot \&
Scott 1998 for more details), sets the
context for our view of the future.  Even at a casual and sceptical
glance these experiments seem to be converging on a power spectrum
which has a peak in it.  This is confirmed by careful quantitative
analysis of combined data sets (Bond, Jaffe \& Knox~1998).
Collectively these CMB measurements already tell us a number of
fundamental things about the sort of Universe that we live in (see
Lawrence, Scott \& White~1999).  The prospects for future measurements
look very bright indeed.  Announcements of the value of $\Omega_0$, for
example, are likely to (continue to) come from experiments carried out
from the best terrestrial sites or suspended from stratospheric
balloons, during the next few years.  However, the full belief of the
community in any detailed cosmological conclusions will and should
await the satellite results.
\begin{figure}
\epsfxsize=13cm \epsfysize=12cm \epsfbox{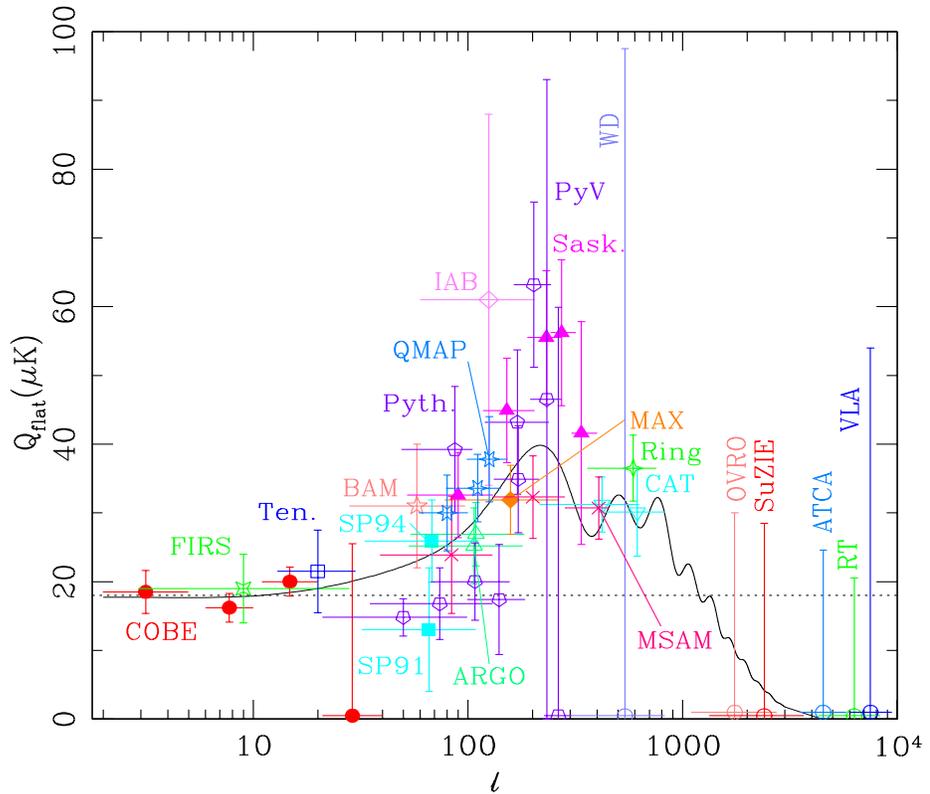}
\caption{A summary of the current anisotropy data, here presented
as the extrapolated value of the quadrupole moment for a flat power
spectrum, plotted  against the multipole moment, which is roughly an
inverse angle, $\ell\simeq135^\circ/\theta$.  The dotted line is the flat
power spectrum fit to {\sl COBE} slone.  The solid line is the power spectrum
for standard Cold Dark Matter, as an example model.  We give a separate
list of references at the end, and there are several other recent reviews
which discuss current experiments in more detail (e.g.~Lawrence 1998).}
\end{figure}

Despite the steadily improving quality of experiments, we believe that
none of the more recent experiments in Figure~1 would have stood as a
convincing {\it discovery\/} of primordial anisotropy had it not been
for {\sl COBE} (this remark certainly applies to our own experiment,
{\sl BAM}, as much as to any other experiment).  What was critical in the
discovery was the understanding of the roles of galactic contributions
and systematic errors, provided by {\sl COBE}'s all-sky coverage and
comparatively stable operating environment.  It was also crucial for
the discovery process that the DMR on {\sl COBE} and the {\sl FIRS} balloon
program showed a consistent fluctuation amplitude {\it and}, later
analysis showed, correlated structures observed at very different
wavelengths.  Many experimenters had reassured themselves by making
plots showing the similarity of the {\sl FIRS} and DMR power spectra, before
the end of the day on which the DMR results were announced.

There is a lesson arising from the history of the measurement of the
intensity spectrum of the CMB which may be useful here.  There were
plenty of experiments prior to 1990 which appeared to have sufficient
sensitivity to perform useful measurements, many of these with no
obvious source of systematic error.\footnote{We will decline to
provide examples here, reminded as we are of Winston Churchill's
failed attempt to maintain parliamentary courtesy when he said that
half of the members present were not asses.}  The successful 1990
experiments (Gush et al.~1990, Mather et al.~1990) were
performed {\it out of the atmosphere\/}, they were {\it
differential\/} and they were carried out with {\it a fanatical
attention to avoiding systematic errors\/} as the primary design
guideline.  The results were clear and reliable enough to render moot
any lingering debates about inconsistencies between previous
experiments.  One should not be surprised to see a very similar
scenario play itself out in the near term anisotropy measurements.
 
\section{Near Term Future Experiments}
 
Table~1 lists the properties of some future experiments.  The list is
meant to be illustrative of current planning; experiments are included
which are past the proposal stage and for which no results are yet
available.  Some of the listed experiments already have data.  Of
course many experiments which have already produced some results and
are therefore not on this list will also produce future results.
All of the listed experiments involve dedicated, custom-built
instrumentation.  The control of systematic errors which this allows
puts these experiments well ahead of attempts to use existing general
purpose facilities.
 
\begin{table}
\caption{Several Future Anisotropy Experiments}
\begin{center}
\begin{tabular}{llll}
\tableline
Snappy &        Frequency &$\ell$-range & Number of \\
Acronym &       Coverage (GHz)& &    freq. channels     \\
\tableline
\multicolumn{4}{r}{Single Dish Telescopes}\\
\tableline
{\sl MAT} &       26--46, 140--150  & 30--850      & 3 \\
{\sl MAXIMA}&     150--420  &   50--700    & 4  \\
{\sl BOOMERanG} & 100--800     & 10--700 &  4 \\
{\sl BEAST  } &       25--90   & 10--800      & \\
{\sl TopHat } & 150--720       &  10--700 & 5  \\
{\sl ACBAR }  &150--450    &     60--2500  & 4  \\
\tableline
\multicolumn{4}{r}{Interferometers}\\
\tableline
{\sl VSA }&   26--36 & 130--1800& 6   \\
{\sl CBI }&   26--36 & 300--3000& 10 \\
{\sl DASI} &  26--36 & 125--700 & 5  \\
{\sl MINT} &  140--250 & 1000--3000 & 10  \\
\tableline
\multicolumn{4}{r}{Satellites}\\
\tableline
{\sl MAP} &   20--106 &  1--800& 5   \\
{\sl Planck} & 30--850 & 1--1500 & 4 + 6  \\
\end{tabular}
\end{center}
\end{table}
 
Sufficient sensitivity is achievable, sometimes through great
technical effort.  The various detection technologies available each
impose constraints on experimental design and come with their own set
of sources of potential systematic errors.  Any serious discussion of
specific systematic errors is beyond the scope of this article but we
include some naive examples to illustrate the problem.  Either a $100\,$mK
change in the temperature of an {\it ideal\/} aluminum mirror or a
$200\,$mK change in the atmosphere above a stratospheric balloon causes a
radiative signal {\it 25 times larger\/} per pixel than the {\sl MAP}
systematic error budget!

\subsection{Systematic Errors}

The careful CMB experimenter is not paranoid, but knows that the
Universe is {\it in fact\/} trying to ruin the experiment.  The standard
answer to the question of what level of systematic error is tolerable
is that there is no systematic way to handle systematic errors and,
therefore, that {\it any\/} level of systematic error is a concern.  We
will ignore this good advice for a moment and try to estimate an
answer.
 
If the goal of an experiment is to get a rough estimate of the power
spectrum of the sky, a systematic error amounting to  $10\%$ of the
signal amplitude contributes about $1\%$ to the power spectrum.  Even
if the signals are correlated in some surprising way  and this
estimate is wrong by a factor of a few, the effect is not likely to
mask the presence of the main acoustic peak, for example.  This fact is what
has allowed us to get so far without a better understanding of diffuse
foreground sources.
 
  On the other hand, there are important questions whose answer requires
correlating many pixels in a map together in order to pick out a fairly weak
efect.  Measuring amplitudes of non-Gaussian statistics of a map or
searching for intensity-polarization correlations are examples.  In
these cases the requirement for what level of signal systematic errors
can contribute to a map becomes very stringent.  The amplitude of
systematic errors should be below the experiment's single pixel
variance divided by the square root of the number of pixels to be
averaged. As a numerical example, in an experiment with $0.13^\circ$
pixels and $30\,\mu$K variance averaging 1/10 of the sky, one needs to
know that systematic errors are less than $0.06\,\mu$K rms for the
average value not to be tainted.  This is 50 times better than any
experiment we have heard of.  The lesson is:
to produce maps of the CMB which merit
careful scrutiny, avoid systematic errors like the
plague.\footnote{Winston Churchill also said `One ought never to turn one's
back on a threatened danger and try to run
away from it.  If you do that, you will double the danger. But if you meet
it promptly and without flinching, you will reduce the danger by half'.}
 
\subsection {Detection Techniques}
Detectors fall into two broad categories: coherent detectors, in which
the radiative electric field, including its phase, is amplified before
detection; and incoherent detectors, which measure total radiative
power within some frequency band.
 
There are two types of very low noise coherent amplifier: InP high
electron-mobility transistors (HEMTs); and
superconductor-insulator-superconductor (SIS) mixers.  HEMTs can be
operated at temperatures as warm as room temperature.  The noise
performance gets better as they are cooled, down to $\simeq4\,$K,
although amplifiers exhibit gain fluctuations at low temperature.
Recently HEMT amplifiers have been made to work at frequencies well
above $100\,$GHz -- noise performance is better at lower frequencies.
SIS mixers are typically quieter than HEMTs and can operate at
frequencies as high as $1\,$THz.  However they must be cooled to
$4\,$K to operate.  Either of these coherent amplifiers can be used in
a single telescope where the signal is amplified and detected, or as
part of an interferometer in which case amplified signals from several
telescopes are each multiplied with a local oscillator signal yielding
lower frequency outputs which are then correlated to produce
interference fringes.
 
The advantages of coherent detectors are that they are fast, stable,
not sensitive to microphonic pick-up and involve simple cryogenics.
HEMTs also have the important practical advantage that many aspects of
detector performance can be verified at room temperature, which
greatly speeds up new instrument development.  The disadvantage is
that they are not as sensitive to broad band signals as incoherent
detectors are.
 
\begin{table}
\caption{Detection strategies}
\begin{center}
\begin{tabular}{llll}
\tableline
Snappy &   Detectors & Striking  & Location \\
Acronym &   & Feature  &   \\
\tableline
\multicolumn{4}{r}{Single Dish Telescopes}\\
\tableline
{\sl MAT   }  &HEMTs and SIS  & Has data & Chile 17{,}000$^\prime$\\
{\sl MAXIMA}&  $100\,$mK Bolos.               &Has data   & Balloon\\
{\sl BOOMERanG} &$300\,$mK Bolos. &First CMB LDB flt.& Balloon,
LDB\tablenotemark{a}\\
{\sl BEAST  } &    & & Balloon, LDB\\
{\sl TopHat }        &     Bolometers  &Tel. {\it above\/} balloon
& Balloon, LDB\\
{\sl ACBAR }  &$300\,$mK Bolos.   & Imaging array  
& S. Pole 10{,}000$^\prime$\\
\tableline
\multicolumn{4}{r}{Interferometers}\\
\tableline
{\sl VSA } &HFETs &14 antennae  & Tenerife\\
{\sl CBI }& HEMTs at $6\,$K   &13 antennae  & Chile, 17{,}000$^\prime$\\
{\sl DASI} &Cooled HEMTs    & 13 elements & S. Pole, 10{,}000$^\prime$\\
{\sl MINT} &SIS   &6 antennae  &Chile\\
\tableline
\multicolumn{4}{r}{Satellites}\\
\tableline
{\sl MAP} & HEMTs  $<95\,$K & Differential tels.
 & Space, L2\tablenotemark{b}\\
{\sl Planck} &HEMTs at $20\,$K  &  & Space, L2\\
    &$0.1\,$K Bolos.   &  &  \\
\end{tabular}
\end{center}
\tablenotetext{a}{LDB = Long Duration Balloon}
\tablenotetext{b}{L2 = Earth-Sun L$_2$ Lagrange point}
\end{table}

Incoherent detectors, in this case bolometers, can be an order of
magnitude more sensitive than HEMT and SIS systems.  They can be made
to operate with background limited performance (BLIP), where
fundamental thermodynamic fluctuations in the incident radiation field
dominate over detector noise.  In addition
they can be made to be sensitive to a
broad range of wavelengths.  However, physical device size scales with
wavelength and so it is easier to make small bolometers sensitive.
Typically bolometers are designed for frequencies above $50\,$GHz.
Bolometers are often susceptible to microphonic and radio-frequency
pick-up.  They are non-linear and therefore they must be characterized
in their experimental operating condition, which can be very difficult
for balloon and satellite experiments.  They need cumbersome
cryogenics to reach their operating temperatures of $0.3\,$K or
colder.  However, their extraordinary sensitivity and broad frequency
coverage often outweigh these disadvantages.  Table~2 lists some
detector properties for the experiments in Table~1.

\subsubsection{Interferometers}
The idea of building a dedicated interferometer to study anisotropy of
the CMB is not new, but improvements in detectors, and especially in
broad bandwidth correlators has made this a very promising option,
which is being actively pursued by several groups.

Interferometers do a good job of rejecting the effects of atmospheric
variations compared to beam-chopped single telescope instruments.
Measurements take place essentially instantaneously, on time scales
associated with the interference bandwidth, and on these time scales
the atmosphere does not vary.  Also, interferometers measure at
slightly higher angular resolution than a single telescope of the same
overall size, and in any case can easily be built for higher
resolution than the currently planned space missions.  This advantage
will be important in exploring the expected Sunyaev-Zel'dovich forest,
especially if they can also be made to work above $200\,$GHz.

\begin{figure}
\epsfxsize=8cm \epsfysize=8cm \epsfbox{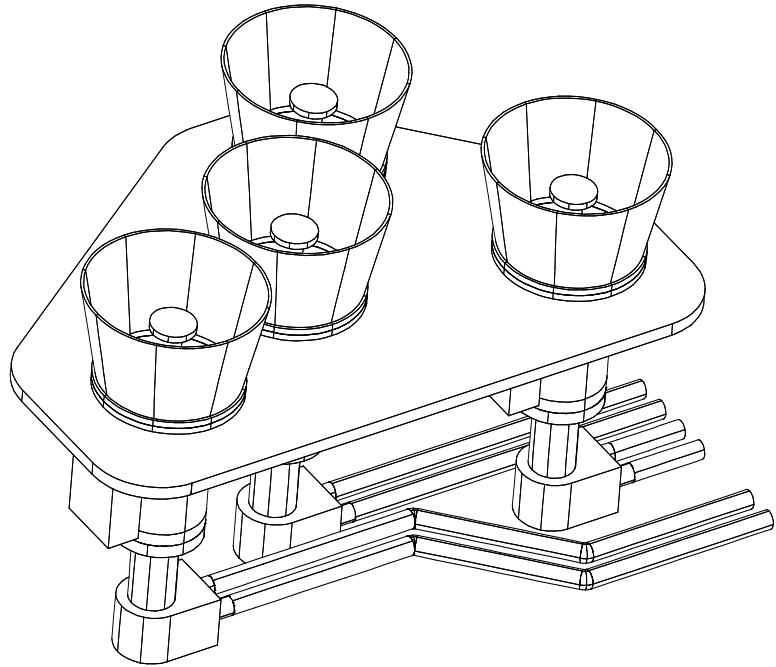}
\caption{The proposed pointed platform of {\sl MINT}, with four $30\,$cm
aperture telescopes mounted on a single $1.5\,$m platform,
illustrating how CMB interferometers are very different from the {\sl
VLA}. Drawing courtesy of W.B. Dorwart, Princeton University.}
\end{figure}

Unlike the case for conventional radio interferometers, the individual
telescopes here are crowded quite close together to keep angular
resolution modest.  Often, all the telescopes are mounted on a single
pointed platform, which eliminates the need for signal delays before
the correlators.  See White et al.~(1997) for an analysis of the performance of
these interferometers for measuring anisotropies.

\subsubsection{Satellites} 
 
Assuming that neither suffers any serious mishap, {\sl MAP} and {\sl Planck}
will produce {\it definitive\/} measurements of the primary anisotropy of the
CMB, at a reliability level which the other experiments can not attain.
The reliability arises form the long observation period, complete sky
coverage and, primarily, the extraordinarily good observing
environment at L2.  Even during the 90-odd day period during which it
makes its way past the moon and out to L2 to start the nominal
observation program, {\sl MAP} will be in a much more thermally and
radiatively stable observing environment than any previous CMB
experiment.

\section{After {\sl MAP} and {\sl Planck}}

What will the important experimental questions be after {\sl MAP} and
{\sl Planck} succeed?  Clearly, measurements of the polarization of
the CMB, which are explored elsewhere in this volume, will be very
exciting.   We also expect that studying diffuse
foreground emission will become very exciting and active, as our ability
to measure and identify these sources of emission develops.  However,
that topic is covered in the whole rest of this book so we need not
consider it further here!  For the remainder of this article we will discuss
various ideas for what might be conceivable in the more long term
future.\footnote{Ignoring the sound advice of Winston Churchill, who said
`It is a mistake to try to look too far ahead. The chain of destiny can only
be grasped one link at a time.'}

\subsection{Anisotropy}

\subsubsection{Statistics}

Can phases contain information which is not {\it more\/} easily seen
in the power spectrum?  In principle, of course the answer is yes.
But in practice, it seems clear that the smart money has to be on the
negative answer.  So although it would always be foolish to neglect to
search for other signals, we expect that the vast majority of the
primary anisotropy information will be contained in the power
spectrum. Partly this is because the signals seem likely to be close
to Gaussian, but also because the power spectrum (or the variance as a
function of scale) is such a robust quantity -- specific patterns
on the sky may require lots of phase-correlation to produce them, but
much of that simply specifies the specific realization, rather than
containing information about the underlying model.  The supremacy of
the power spectrum will certainly cease to be true for foreground
signals, or indeed for a range of astrophysical processing effects
that come in at smaller angular scales.

One could imagine mounting a specific search for, e.g.~point or line
sources on the sky, as specific examples of non-Gaussian signals.  One
question to ask, then, is what sort of strategy one would design to carry
this out most efficiently ({\it and\/} convincingly).  We find it hard to
see how to avoid the conclusion that you would end up making a map, perhaps
deeper and with higher resolution than otherwise, but a map nevertheless.
Hence we suspect that the search for non-Gaussian signals is unlikely to
be a strong driver for the design of future experiments, even if it plays
a stronger role in the data analysis.

A great deal of effort has been going into the study of non-Gaussian
signals, e.g.~using Minkowski functionals, wavelets, etc.
Given how many such tests have
already been applied to {\sl COBE}, we imagine that {\it every\/} reasonable
statistic will be measured for all future large data-sets.  In particular
we foresee an increased interest in the investigation of non-Gaussian
statistics for various sorts of {\it foreground\/} signal.

\subsubsection{Angular scales}
Ignoring foregrounds, how far out in $\ell$ is far enough?
{\sl Planck} seems sufficient for the primary signal.  But that may change,
depending on what we learn about foregrounds and the secondary signals,
caused by various astrophysical effects, which conceptually lie in the
`grey-area' between background and foregrounds.  There seems to be a growing
interest in these astrophysical signals at small scales, and we see no
reason for that to change.  It may be that the smallest
angular scales are ultimately best probed with interferometers.
We expect there to be secondary signal information down to the angular
scales of distant galaxies, i.e.  $\ell\simeq{\rm few}\times10^4$.
         
\subsubsection{A CMB Deep Field}
What might we learn from a CMB deep field?  Of course, irrespective of
the answer to that question, it will be done anyway!  Non-Gaussian
signals from higher-order effects at small-scales would certainly show
up in such a map.  On scales where there has been significant growth
of structure, and certainly on non-linear scales, we would expect
there to be significant non-Gaussianity.  There seems little doubt
that at some point, when the instrumentation has matured, it will be
worthwhile to carry out such a CMB Deep Field.  Exactly how
non-Gaussian (or in other ways surprising) the small-scale signals
turn out to be will determine how far beyond `cosmic variance' it is
worth integrating.

\subsubsection{Sunyaev-Zel'dovich effects}
We are sure that, motivated by the impressive results of today's
experimenters, investigation of S-Z effects will continue to grow as a
sub-field.  Particularly exciting is the idea of `blank sky' searches for
the `S-Z Forest', or ionized gas tracing out the Cosmic Web.

The thermal Sunyaev-Zel'dovich effect, or inverse Compton scattering of the
CMB photons through hot gas, gives a temperature fluctuation of
roughly $(kT_{\rm e}/m_{\rm e}c^2)\tau$, where $T_{\rm}$ and $m_{\rm e}$ are
the temperature and mass of the electron, respectively, and $\tau$ is
the optical depth through the ionized gas.  There is also a spectral shape,
distinct from the CMB blackbody, of a well-known form (see e.g.~Sunyaev \&
Zel'dovich~1980)
Detailed studies of the thermal S-Z effect for particular clusters will
provide constraints on the morphology, clumping, thermal state of the gas
and projected mass, amongst other things.  The power spectrum of these
fluctuations peaks at $\ell\simeq2000$ typically, with 
$\left\langle Q_{\rm flat}\right\rangle$ amplitude
of a few $\mu$K, although with great variation between models.  Detailed
investigation of this power spectrum might further distinguish between
cosmological models, and between ideas for cluster formation.  The power
spectrum for the kinematic effect, and for related effects (due to variations
in potential, for example) are generally much smaller.

Several of these
`higher-order' Sunyaev-Zel'dovich type effects are potentially
measurable for {\it individual\/} clusters,
and will doubtless be attempted in the future (see the review
by Birkinshaw~1998).  Certainly the kinematic effect (which depends on
the line-of-sight velocity and is $\sim(v_{\parallel}/c)\tau$)
can be measured for some clusters.  However, this
effect has the same spectrum as a CMB fluctuation, and so the small-angle
CMB anisotropies act as a source of `noise', making is difficult to
measure the velocities to better than a few hundred ${\rm km}\,{\rm s}^{-1}$.
One further effect uses the polarization in the CMB scattered by the
kinematic S-Z effect, which depends on the cluster's transverse velocity
(actually $\sim (v^2/c^2)\tau$).
In principle, together with the kinematic S-Z effect itself, this gives a
means of estimating the full 3 dimensional velocity of clusters.  Although
difficult to measure, this polarization signal has a frequency dependence
which may help to disentangle it from other effects (Audit \& Simmons~1999).
There are other spectral signals expected from non-thermal electron
populations, for example in the lobes of radio sources.  However,
the utilisation of
such measurements to study the lobe properties will require extremely high
angular resolution.

\subsubsection{Other secondary effects}
There are several other known secondary effects (see, e.g.~Hu et al.~1995,
and other contributions to this volume), and surely many other
{\it unknown\/} ones!

One effect which has been studied in some detail is a second-order
coupling between density and velocity, usually referred to as the Vishniac
effect.  In a sense this can be thought of as specific case of the
kinematic Sunyaev-Zel'dovich effect.  For Cold Dark Matter type models the
signal is typically $\simeq1\,\mu$K and peaking at 
$\ell\simeq{\rm few}\times10^3$ (e.g.~Hu \& White~1996,
Jaffe \& Kamionkowski~1998).  Although
certainly difficult to measure, it is nevertheless feasible, and worth
pursuing, since measurement can give direct information about reionization.
Additional structure ar small scales (e.g.~from an isocurvature mode) could
increase the signal.  In addition there will be polarization effects,
although these are likely to be {\it really\/} small.

Patchy reionization (discussed elsewhere in this volume) is just another
S-Z effect, and tends to be dominated by the kinematic source from moving
bubbles of gas as the Universe undergoes reionization.  The amplitude of
this signal seems likely to be smaller than for the Vishniac effect,
although it is as yet
unclear what the predictions will be for realistic models which include
inhomogeneous reionization (with radiative transfer through
voids etc.), distributions of sources, and other complications.  Again
there may be polarization effects, and correlations with other signals,
which, in principle, could be used to pull the signal out.
In addition there may also be a measurable S-Z
signal from the Ly$\,\alpha$ forest, on
scales well below an arcminute, and with amplitude perhaps as high
as a few $\mu$K (Loeb~1996).

Rees-Sciama, or varying potential fluctuations tend to be rather
small in amplitude ($<10^{-6}$ in fractional temperature change), but
not negligibly so.  Here again there are a number of effects, in particular
those caused by time-varying potentials in the light-crossing time, and
those caused by potentials moving across the line of sight (e.g.~Tuluie,
Laguna \& Anninos~1996).  These will have
CMB-like spectra, and the signal will be dominated by non-linear
structures (meaning that the statistics will be highly non-Gaussian).
The effects may peak at relatively small angles $\ell\simeq{\rm few}\times100$,
but there they will be well below the primary signal, and hard to
disentangle.  So the best prospects for detection may be at smaller scales,
where the primary power spectrum is falling off.  Detection may also be
easier using correlations with other signals.  And certainly such signals
are unlikely to be Gaussian, and so may be teased out of the data by looking
at their statistics (e.g.~Spergel \& Goldberg~1999).

Gravitational lensing affects the CMB power spectrum by smearing the
anisotropies, thereby smoothing out features in the power spectrum.  The
temperature field is affected by this smearing, so that combinations of
derivatives can be used to extract the lensing signal directly, at least
in principle (Seljak \& Zaldarriaga~1999).  
The projected matter field can be reconstructed through a combination of
this technique and correlations with other signals (Zaldarriaga \&
Seljak~1999).  For example, the large angle signal caused by the variation
in gravitational potential (the `ISW effect') may be correlated with the
lensing signal in open or $\Lambda$-dominated models.   However, the level
of such a correlated signal is not likely to be large.  One can easily
imagine searching for all sorts of other correlations, for example the
lensing signal with S-Z signals, with surveys
at other wavelengths, e.g.~large-scale structure, X-ray maps, etc.

\subsubsection{Spatial-spectral signals}
At the moment the only significant signal which mixes both spatial and
spectral deviations is the S-Z effect.  Although we have no specific
ideas, we imagine that other such effects, involving perhaps different
scattering processes, are likely to be developed in the near future.
Although we expect the effect to be quite small, we mention as an
example that Rayleigh scattering, which would spectrally filter
anisotropy signals, has been omitted from calculations.  In addition there
could in principle be resonant line scattering from molecules in clouds
at high redshift.  Searches for
such mixed spatial-spectral signals seem likely to become more
important as multi-frequency data-sets improve in quality and quantity. 

\subsection{Non-anisotropy}

Non-anisotropy measurements are heroically hard to do; certainly such things
are worth pursuing, but the immediate returns are not as obvious as for
the current anisotropy prospects.  On the other hand, we expect that effort
will fairly soon return to this direction when the `easy' results have been
mined from the primary power spectrum.  Here we simply list a number of
possibilities.  Figure~3 shows the form of some of the standard distortions
to the CMB spectrum.
\begin{figure}
\epsfxsize=11cm \epsfbox{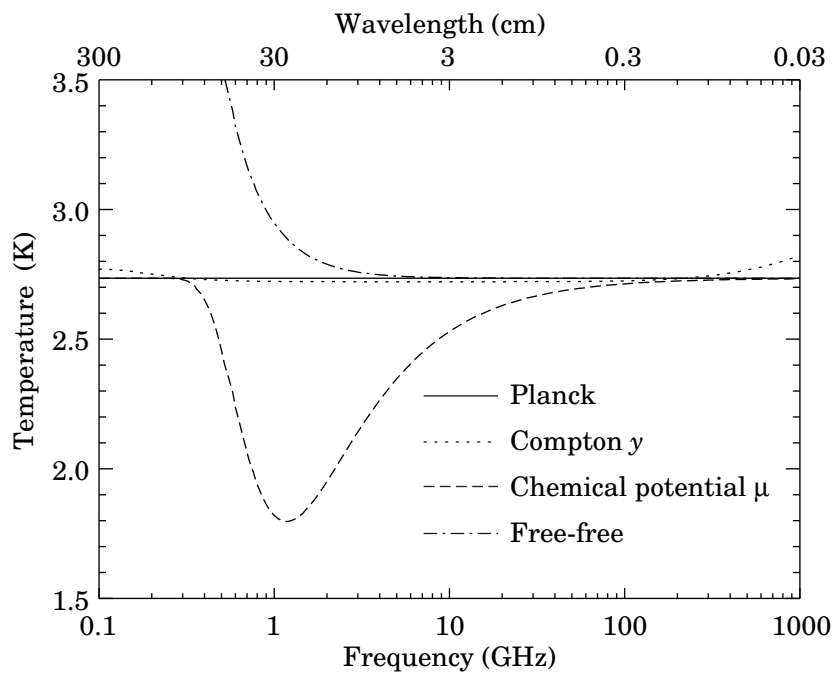}
\caption{Shapes for some theoretical possibilities for spectral distortions.
The amplitudes here are arbitrary. However, the FIRAS 95\% confidence
limits for the amplitudes of these distortions are that the chemical
potential is less than $15\,$mK at the peak and that free-free is less
than $10\,$mK at $4\,$GHz.  At those frequencies the Galaxy is perhaps
100 times brighter.  See Smoot (1997) for further discussion.}
\end{figure}

A FIRAS/COBRA style total emission measurement of the spectrum of
the sky will almost entirely be dominated by foregrounds outside of
the 20 to $400\,$GHz frequency window in which the spectrum is already
well measured.  The present FIRAS limits allow parametrized spectral
distortions as large as 10s of mK, easily larger than the
measurement uncertainty in a careful experiment, but 100 to 1000 times dimmer
than diffuse galactic emission at those wavelengths! Perhaps a
multi-frequency measurement with appropriate angular resolution and
sky coverage will allow a reliable extrapolation to zero galactic
emission, but it will not be easy. Details for commonly considered
distortions are listed below.

\subsubsection{Compton distortions}
$y$-distortions have essentially already been measured, in the
sense that the sum of all the S-Z detected structure will give the
uniform Compton-distortion over the sky.  Certainly this gives a lower limit,
which seems likely to be the bulk of the detectable signal (barring
unforeseen exotic processes).  The size of this signal is
estimated to be $y\sim10^{-6}$
(e.g.~Colafrancesco et al.~1997), depending on the cosmology.
After the {\sl Planck}
mission (and S-Z investigations from ground-based interferometers) we will
have a very precise estimate for the uniform $y$-distortion (and indeed
an estimate of its power spectrum as well).  Between this underlying
signal, and the FIRAS upper limit on a full-sky distortion, there will
remain only a rather narrow window to search for possible isotropic
$y$-distortions from other sources (such as late energy injection, unrelated
to cluster formation).  Since there are no immediate candidates for such
processes, and the window is rather narrow, we don't see this as a
particularly strong motivation for mounting a next generation FIRAS mission.

\subsubsection{Free-free emission}
For late energy releases, free-free emission leads to a distortion in
the CMB spectrum, which increases towards lower frequency.
This seems to be the type of distortion which is most feasible to measure
in the near future for realistic models of the Universe.  The best upper
limits at the moment imply free-free optical depths of order
$Y_{\rm ff}\la10^{-5}$ (e.g.~Nordberg \& Smoot~1998).
Since this distortion increases at lower frequencies, then it is best
investigated at the lowest frequency at which foreground signals can be
dealt with, which means somewhere around $5\,$GeV.  The expected signal
at these frequencies may be as high as $100\,\mu$K,  corresponding to
$Y_{\rm ff}$ only about an order of magnitude below the current limits.
The planned experiments ARCADE and DIMES (Kogut~1996)
may be able to reach into the
parameter space for realistic models, and help us understand more about
the early ionized stages of the intergalactic medium.  One nice thing
about free-free is that lowering the temperature of the ionized medium
{\it increases\/} the distortion (approximately
$\propto n_{\rm e}^2/\sqrt{T_{\rm e}}$), even although it decreases the
Compton distortion.  Hence good limits on $Y_{\rm ff}$ imply either low
reionization redshifts or high electron temperatures, and limits on
$y$ would restrict $T_{\rm e}$, so that direct limits on $z_{\rm reion}$
could be obtained.

\subsubsection{Chemical potential}
Current limits on $\mu$-type distortions are at the $10^{-5}$ level.
Note that this allows
about $15\,$mK at $1\,$GHz within the error budget of the
measurements, which is about 0.1\% of the galactic signal.  So pushing that
limit further down is going to be tricky!  The way to do this would
presumably be to make a spectral map of the sky and extrapolate to zero
galaxy (essentially what FIRAS did).  So how big could a signal be?

Some amount of $\mu$-distortion is unavoidable, since it is generated by
the damping of small-scale perturbations.  For realistic models the value
is likely to be around ${\rm few}\times10^{-8}$ (Hu, Scott \& Silk~1994),
which seems unlikely ever to be measurable.  Of course various exotic
processes, including energy injection at redshifts $z\sim10^5$ could give
much higher values of $\mu$.  Limits could be set by experiments which
also constrain free-free signals.  However, we see no compelling reason
currently to invest heavily in future experiments seeking to measure $\mu$
itself.  Of course, if any hint of signal were to turn up then that would be
extremely exciting (since unexpected) -- in that case further investigation
of the turn-off in the distortion at low frequencies would probe an
otherwise unexplored early epoch.

\subsubsection{Recombination lines}
When the Universe recombined, every atom emitted at least one Lyman
photon, or else got from the first excited state to the ground state
via the two-photon process (see Seager, Scott \& Sasselov~1999 for more
details).  This is a lot of photons, waiting there
to be discovered!  Mere measurement of the background of these photons
would be an unprecedented confirmation of the Big Bang paradigm, that the
Universe became neutral at $z\simeq1000$.  Further investigation of these
recombination lines would be a direct probe of the recombination process,
and might provide further cosmological constraints.  For example, the
strengths of the residual Ly$\,\alpha$ feature and the two-photon
feature will depend on the baryon density and on the expansion rate, hence
allowing measurements of $\Omega_{\rm B}$ and $H$ at $z\simeq1000$.

The problem is that the main recombination lines lie at wavelengths
$\lambda\simeq150\,\mu$m, where the signal from the galaxy is orders of
magnitude stronger.  The way to try to find the signal is then presumably
to have enough spatial information to be able to extrapolate to zero
Galaxy, and at the same time to have adequate spectral information to
distinguish the relatively wide spectral feature.  If all else failed
it might be possible to rely on the dipole to extract the cosmological
signal, but that would be even more difficult.  So we might envisage an
experiment with reasonable sky coverage, low angular resolution, but at
least 3 spectral channels (say in the range 100--200$\,\mu$m) to extract
the wide line.  The spectral resolution would have to be good enough to
distinguish this from a roughly isotropic component of warm interstellar
dust -- but that should be possible given that the spectral shape of the
recombination lines is calculable (Dell'Antonio \& Rybicki~1993,
Boschan \& Biltzinger~1998).
  
\subsubsection{21\,cm line studies}
If the Universe became reionized at redshifts between 5 and 20 there
should be a spectral feature due to red-shifted $21\,$cm emission from
neutral hydrogen which appears today at $70$ to $240\,$MHz (see
Shaver et al.~1999).  This emission can be seen against the CMB provided that
there are spatial or spectral signatures (e.g.~Tozzi, et al.~1999) and
a mechanism which decoupled the electron spin temperature from the
CMB.  In principle, such studies, using the proposed Square
Kilometer Array for example, could provide information about the
processes that marked the end of the so-called Dark Ages, i.e.~the
reionization process and the formation of the first structures.
This endeavor is sometimes called `cosmic tomography'.

\subsubsection{Other diagnostics of the `Dark Ages'}
There are of course other ways of probing the end of the Dark Ages, and
even into that epoch, many of which might come from entirely different
wavelengths, for example the near infra-red with {\sl NGST}.  However, we
imagine that the microwave band will continue to be important in furnishing
new ideas for exploring the domain between $z=5$ and $z=1000$.  One recent
speculative idea involves searching for masers which may come from structures
at either the recombination or reionization epochs (Spaans \& Norman~1997).
There will surely be other such ideas in the coming years.

\subsubsection{Measurements of $T_{\rm CMB}(z)$}
The currently best value for the CMB temperature is
$T_0=2.725\pm0.001\,$K (Mather et al.~1999).  It seems unclear why
anyone should care about a more precise measurement than that!  Before
the existence of the CMB was even suspected, there was evidence for
excess excitation in line ratios of certain molecules, notably
cyanogen , in the interstellar medium (McKellar,~1941).
This method has more recently been used to
constrain the CMB temperature at high redshifts ($z\sim{\rm few}$)
using line with excitation temperatures of the relevant energy
(e.g.~Songaila et al.~1994, Roth \& Bauer~1999).  Measurements of
other line ratios etc.~can be used to set limits on the variation of
fundamental physical constants (e.g.~Webb et al.~1999).  In a similar
way, detailed measurement of the blackbody shape indicates that
certain combinations of fundamental constants have not varied much
since $z\sim1000$.
\begin{figure}
\epsfxsize=14cm \epsfbox{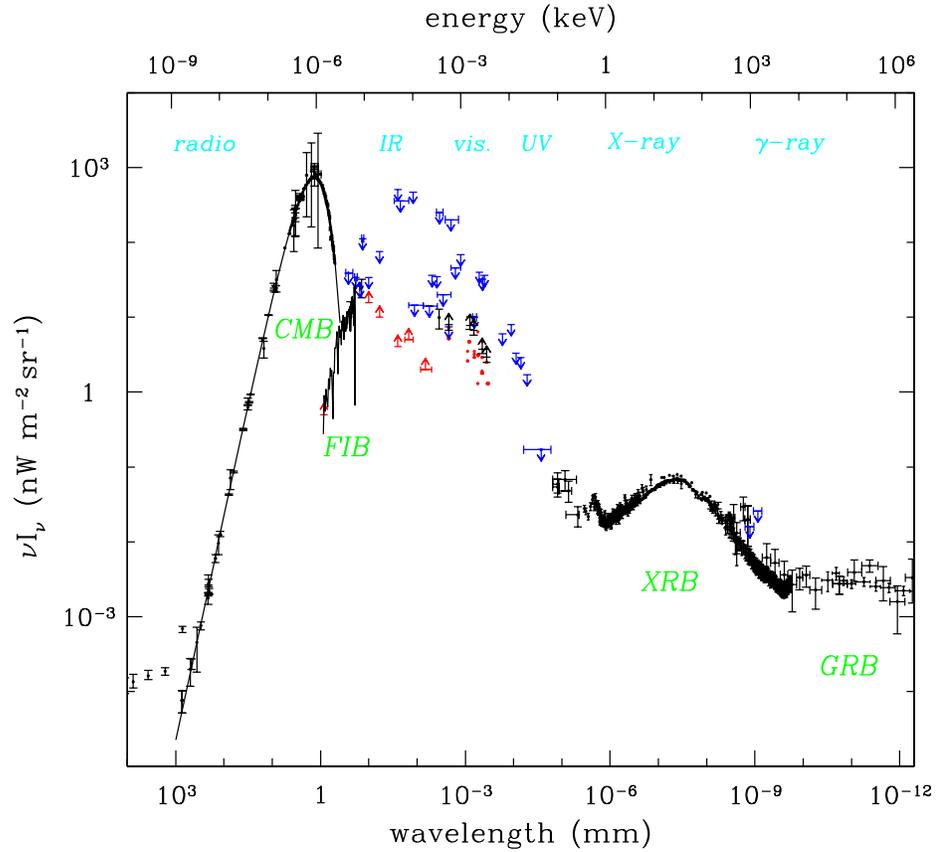}
\caption{A compilation of recent constraints on extragalactic diffuse
background radiation.  In terms of total energy the CMB and Far-Infrared
Backgrounds dominate.  The data are collected primarily from Ressel \&
Turner~1990, Smoot~1997, Lagache et al.~1999, Hauser et
al.~1998, Dwek \& Arendt~1998, Pozzetti et al.~1998, Leinert
et al.~1998, Miyaji et al.~1998, Sreekumar et al.~1998,  and
Kappadath et al.~1999.  In this colour version lower limits are shown in
red and upper limits in blue.}
\end{figure}

\section{Epilogue}

Assuming that {\sl MAP} and {\sl Planck} are fully successful, and that the
current suite of ground-  and balloon-based experiments also return
exquisite data, what then?  Will this be then end of the study of the
CMB?\footnote{Churchill warned that `success is never final'.  He also
pointed out that `it is a good thing for an uneducated man to read books of
quotations'.}
Eventually we can imagine a time when the primordial anisotropies
have been measured so accurately that there are diminishing returns from
further generations of satellite missions, and when small scale
measurements, involving non-Gaussian signals, mixed spatial-spectral signals,
and other complications, have moved firmly into the regime of `messy
astrophysics'.  However, there will be further primordial information to
unlock from ever more ambitious polarization experiments.  Certainly the
CMB should not be looked at in isolation -- although it is the dominant
diffuse extragalactic background, there are several others to study
(see Figure~4).  And if that
doesn't leave the future still filled with exciting and challenging
possibilities, there's always the cosmic neutrino background!

\acknowledgments
We thank the editors for their patience.

\end{document}